\documentclass[aps,pre,floats,twocolumn,showpacs,superscriptaddress]{revtex4}

\usepackage{epsfig}
\usepackage{latexsym,amsmath}

\begin{document}
\title{Tuning clustering in random networks with arbitrary degree distributions}

\author{M. \'Angeles Serrano}

\affiliation{School of Informatics, Indiana University,\\ Eigenmann Hall, 1900 East Tenth Street, Bloomington, IN 47406}

\author{Mari{\'a}n Bogu{\~n}{\'a}}

\affiliation{Departament de F{\'\i}sica Fonamental, Universitat de
  Barcelona,\\ Mart\'{\i} i Franqu\`es 1, 08028 Barcelona, Spain}

\date{\today}

\begin{abstract}
We present a generator of random networks where both the
degree-dependent clustering coefficient and the degree distribution
are tunable. Following the same philosophy as in the configuration
model, the degree distribution and the clustering coefficient for
each class of nodes of degree $k$ are fixed {\it ad hoc} and {\it a
priori}. The algorithm generates corresponding topologies by applying first a closure of triangles and secondly
the classical closure of remaining free stubs. The procedure unveils
an universal relation among clustering and degree-degree correlations for all networks, where the level of assortativity establishes an upper
limit to the level of clustering. Maximum assortativity
ensures no restriction on the decay of the clustering coefficient
whereas disassortativity sets a stronger constraint on its behavior.
Correlation measures in real networks are seen to observe this
structural bound.
\end{abstract}


\maketitle

\section{Introduction}
Models in complex networks science aim to reproduce some common
empirical statistical features observed across many different real
systems, from the Internet to society \cite{Barabasi02,Dorogovtsev02,Newman03}. Many of those models are able
to recreate prominent recurrent attributes, such as the small-world
property and scale-free degree distributions with
characteristic exponents between $2$ and $3$ as measured for
networks in the real world. Other characteristics, such as the
presence, the shape, and the intensity of correlations, are also
unavoidable in models intending to help us to understand how these
complex systems self-organize and evolve.

The first reference to correlations in networks appearing in the
literature is the clustering coefficient \cite{watts98}, which
refers correlations among three vertices. The clustering is a
measure of transitivity which quantifies the likelihood that two
neighbors of a vertex are neighbors themselves. Then, it is a
measure of the number of triangles present in a graph. In addition
to the empirical evidence that the vast majority of real networks
display a high density of triangles, the concept of
clustering is also relevant due to the fact that triangles are
--together with edges-- the most common building blocks taking part
in more complex but elementary recurring subgraphs, the so-called
motifs \cite{Motifs}. It has been argued that networks large-scale
topological organization is closely related to their local motifs
structure \cite{MotifsBara} so that these subgraphs could be related
to the functionality of the network and can be fundamental in
determining its community structure \cite{CommParisi,CommVicsek}.

All these mean that a correct quantification and modeling of the
clustering properties of networks is a matter of great
importance. However, most modeling efforts beyond the degree
distribution have focused in the reproduction of two point
correlations patterns, typified by the average nearest neighbors
degree \cite{knnRomu}, so that clustering is just obtained as a byproduct. In most synthetic networks, it vanishes in the
thermodynamic limit, but, as to many other respects, scale-free networks
with divergent second moment stand as a special case. The decay of
their clustering with the increase of the network size is so slow
that relatively large networks with an appreciable high cohesiveness
can be obtained \cite{Catanzaro05}. Nevertheless, it remains to be an indirect effect and no control over its intensity or shape is practicable. Therefore, an independent modeling of clustering is required and a few growing linear preferential
attachment mechanism have been suggested. One of the proposed models
\cite{ModelClusDoro} reproduces a large clustering coefficient by
adding nodes which connect to the two extremities of a randomly
chosen network edge, thus forming a triangle. The resulting network
has the power-law degree distribution of the Barab\'asi-Albert model
$P(k) \sim k^{-3}$, with $\langle k \rangle=4$, and since each new vertex induces the
creation of at least one triangle, the model generate networks with
finite clustering coefficient. A generalization on this model
\cite{BarratGrow} which allows to tune the average degree to
$\langle k\rangle=2m$, with $m$ an even integer, considers new nodes connected to
the ends of $m/2$ randomly selected edges. Two vertices and three
vertices correlations can be calculated analytically through a rate
equation formalism. The clustering spectrum is here finite in the
infinite size limit and scales as $k^{-1}$. 

Those models do not allow much freedom in the form of the resulting clustering
coefficient, neither in the ensuing degree distribution, so that,
although a valuable first approach, they constitute a timid attempt
as clustering generators. In this paper, we make headway by introducing a generator of random
networks where both the degree-dependent clustering coefficient and
the degree distribution are tunable. After a brief review of several
clustering measures in section II, the algorithm is
presented in section III. In section IV, we check the validity of the algorithm using numerical simulations. Section V is devoted to the theoretical explanation of the constraints that degree-degree correlations impose in the clustering. We find that assortativity allows higher levels of clustering, whereas disassortativity imposes tighter bounds. As a particular case, we analyze this effect for the class of scale-free networks. We end the section by examining some empirical networks, finding a good agreement with our calculations. Finally, conclusions are drawn in section VI.

\section{Measures of clustering}
Several alternative definitions have been proposed over time to quantify clustering in networks. The simplest measure is defined as
\cite{clusteringBarrat,clusteringNewman}
\begin{equation}
C_{\Delta}=\frac{\mbox{3 } \times \mbox{ Number of triangles}}{\mbox{Number
of connected triples}}.
\end{equation}
This scalar quantity does not give much information about local
properties of different vertices because it just counts the overall number of triangles regardless of how these triangles are placed among the different vertices of the network. 

The clustering coefficient, first introduced by Watts and Strogatz \cite{watts98}, provides instead
local information and is calculated as
\begin{equation}
c_i=\frac{2 T_i}{k_i (k_{i} -1)},
\end{equation}
where $T_i$ is the number of triangles passing through vertex $i$
and $k_{i}$ is its degree. The average of the local clustering
coefficients over the set of vertices of the network, $C$, is
usually known in the literature as the clustering coefficient. Watts
and Strogatz were also the first in pointing out that real networks
display a level of clustering typically much larger than in a
classical random network of comparable size, $C_{rand}=\langle k \rangle /N$,
with $\langle k \rangle$ the average degree and $N$ the number of nodes in the
network. Although $C$ and $C_{\Delta}$ are sometimes taken as equivalent, they may be very different, even
though both measures are defined in the interval $[0,1]$. 

With the definition of $c_{i}$ we have gone to the other extreme of
the spectrum --from a global to a purely local perspective-- so that we have highly detailed information. One can adopt a compromise between the global property defined by $C$, or $C_{\Delta}$, and the full local information given by
$c_{i}$ by defining an average of $c_{i}$ over the set of vertices
of a given degree class \cite{DefclusteringkVespas}, that is,
\begin{equation}
\bar{c}(k)=\frac{1}{N_k} \sum_{i \in \Upsilon(k)} c_{i}=\frac{1}{k
(k-1)N_k} \sum_{i \in \Upsilon(k)} 2T_{i} ,
\label{c(k)}
\end{equation}
where $N_{k}$ is the number of vertices of degree $k$ and
$\Upsilon(k)$ is the set of such vertices. The corresponding scalar
measure is called the mean clustering coefficient and can be computed on the
basis of the degree distribution $P(k)$ as
\begin{equation}
  \bar{c} = \sum_{k} P(k) \bar{c}(k),
  \label{eq:30}
\end{equation}
which must not be confused with the clustering coefficient
$C=\bar{c}/(1-P(0)-P(1))$. In fact, we have implicitly assumed that $\bar{c}(k=0)=\bar{c}(k=1)=0$ whereas in the definition of $C$ we only consider an average over the set of vertices with degree $k > 1$. this fact explains the difference between both measures.

In the case of uncorrelated networks, $\bar{c}(k)$ is independent of
$k$. Furthermore, all the measures collapse and reduce to $C$
\cite{clusunNewman,Hiddenv,ClusteringBurda}.
\begin{equation}
  \bar{c}(k) = C_{\Delta}= C = \frac{1}{N}\frac{(<k^2>-<k>)^2}{<k>^3} \mbox{  ,  } k > 1.
  \label{eq:32}
\end{equation}
Therefore, a functional dependence of $\bar{c}(k)$ on the
degree can be attributed to the presence of correlations. Indeed, it has been observed that $\bar{c}(k)$
exhibits a power-law behavior $\bar{c}(k)\sim k^{-\alpha}$ (typically $0\le
\alpha \le 1$) for several real scale-free networks. Hence,
the degree dependent clustering coefficient has been proposed as a measure of hierarchical
organization and modularity in complex networks
\cite{ExampleclusteringBara}.

Recently, a new local clustering coefficient has been proposed, which filters
out the bias that degree-degree correlations can induce on that measure
\cite{ClusteringVazquez}
\begin{equation}
  \tilde{c}_i =  \frac{T_i}{\omega_i},
  \label{eq:33}
\end{equation}
where $\omega_i$ is the maximum number of edges that can be drawn
among the $k_i$ neighbors of vertex $i$. This new measure does not
strongly depend on the vertex degree, remaining constant or
decreasing logarithmically with the increase of $k$ when computed
for several real networks.

\section{The algorithm}

In this paper, we develop and test a new algorithm that, in the same
philosophy of the classical configuration model (CM), generates networks with a given
degree distribution and a preassigned degree
dependent clustering coefficient $\bar{c}(k)$, as defined in Eq. (\ref{c(k)}). 

The CM has been one of the most successful
algorithms proposed for network formation \cite{Molloy95,Molloy98}. The relevance of the algorithm
relies on its ability to generate random networks with a preassigned
degree sequence --taken from a given degree distribution-- at the user's discretion
while maximizing the network's randomness at all other respects. The
algorithm became relevant as soon as more real networks were
analyzed and proved to strongly deviate from the {\it supposed}
Poisson degree distribution predicted by the classical model of
Erd\"os and R\'enyi \cite{Erdos59,Erdos60}. Ever since, the CM has been extensively used as a
null model in contraposition to real networks with the same degree
distribution.

One of the well-known properties of the CM is that clustering
vanishes in the limit of very large networks (see Eq. (\ref{eq:32})) and, thus,
it clearly deviates from real networks, for which clustering is
always present. In general, a high level of clustering may change
the percolation properties of the network, alter its resilience in
front of removal of its constituents, or affect the dynamics that
take place on top of them. Since such processes inextricably entangle topology and functionality, it would be
very interesting to have at one's disposal an algorithm that
generates clustered networks in a controlled way so that one can
check which is the real effect of transitivity on its topological
and dynamical properties.

With this purpose, we introduce an undirected unweighted static model where the total
number of nodes in the network, $N$, remains constant, as in the case of the CM. The algorithm
comprises three different parts: 1) Assignment of a degree to each
node and assignment of a number of triangles to each degree class
according to the expected distributions. 2) Closure of triangles
and 3) closure of the remaining free stubs. In what follows, we give a detailed description of the algorithm.\\

\noindent
{\bf 1) Degree and clustering from expected distributions}
\begin{itemize}
\item An {\it a priori} degree sequence is chosen according to a
given distribution $P(k)$, so that each vertex is awarded an {\it a
priori} number of connections in the form of a certain number of
stubs.

\item An {\it a priori} clustering coefficient $\bar{c}(k)$ is also fixed, so
that each class of nodes of degree $k$ is assigned an {\it a priori}
number of triangles \footnote{The expected number of triangles associated to each class is
$T(k)=\frac{1}{2}k(k-1)\bar{c}(k)P(k)N$. For very large degrees, it may happen that this number is smaller
than one. To overcome this problem, one can determine $T_{k}$ by allocating a number of
triangles, $T_{i}$, to each vertex $i$ using, for
instance, the binomial distribution $B(p,N_{max})$ with $p=\bar{c}(k_{i})$ and $N_{max}=k_{i}(k_{i}-1)/2$. Then, the number of triangles $T_{i}$ are summed up for each class. Once each class has been assigned a total number of triangles, individual vertices forget
the initially ascribed $T_{i}$.}. Note that the number of triangles is fixed for
the whole class and not for the particular vertices of the class.
This is a key point of the algorithm because fixing the number of
triangles to each single vertex would impose a number of constraints
that would make nearly impossible to close the network.

\item All nodes begin with a number of $0$ associated edges and
all degree classes begin with a number of $0$ associated triangles.
\end{itemize}

\noindent
{\bf 2) Triangle formation}\\

First some preliminary remarks. Both stubs and edges can be selected to
form a triangle. Stubs are half links associated to one node, and
edges are entire links associated to two nodes, and thus have double
probability with respect to stubs to be selected to participate in a
triangle. Let us define the set of eligible components (EC) as the set
of free stubs and edges associated to nodes belonging to degree classes with a
number of triangles below its expected value (unsatisfied classes).
Stubs and edges of nodes in satisfied classes should not be in the set of EC. Edges of nodes which cannot form more triangles (with only
edges as components and neighbors without stubs) should not be in the set of EC. Stubs and edges of nodes with only one component should not be in the set of EC. Notice that the set of EC changes dynamically as triangles are formed.

The algorithm then proceeds by choosing three different nodes and forming a triangle among them whenever it is possible and it did not exist previously. The selection of the nodes is performed hierarchically as follows.
\begin{itemize}
\item For the first node, a degree class $k_{1}$ among the ones with
unsatisfied number of expected triangles is chosen with a certain
probability distribution, $\Pi(k)$, not necessarily uniform (the specific form for this function and the motivation to introduce it will be discussed at the end of the section and a theoretical explanation is given in section \ref{theoretical}). Then, the node is selected through a
component which is chosen with uniform probability within the subset
of eligible components in the class, EC$(k_{1})$. A second different component of the same node is
selected.
\item If the two chosen components are edges,
and the second and the third nodes at the end of the edges still
have free stubs, the triangle is formed by merging one free stub of
the second node and one free stub of the third node (see Fig. \ref{figD1}).
\begin{figure}[h]
\epsfig{file=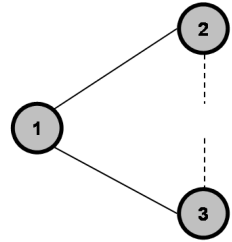,width=3.5cm}
 \caption{The two selected components of the first node,
  marked 1, are edges. The triangle if formed by connecting nodes 2 and 3, whenever they have free stubs.} \label{figD1}
\end{figure}

\item If one component is a stub and the other an edge, a
third node is necessary. First, a new component is selected for the
second node at the end of the edge. If it is an edge, the triangle is formed
by merging one free stub of the first node and one free stub of the
third node. If it is a stub, then a third node is chosen in the same way as the first one under the condition that it has two free stubs. The triangle is then formed by merging these two free stubs with the ones of the first and second nodes (see Fig. \ref{figD2}). 
\begin{figure}[h]
\epsfig{file=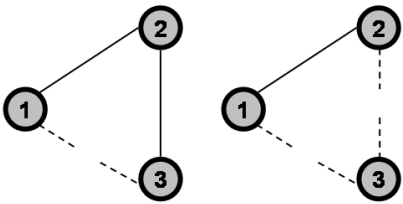,width=6cm}
 \caption{The two selected components of the first node are one edge and one stub. A second component is chosen for the second node. On the left side of the figure, the component is an edge whereas on the right side it is a free stub.} \label{figD2}
\end{figure}

\item It may happen that the two components of the first node
are stubs. Then, a new node is selected in the same way as the first one under the condition of having at least one free stub, and a
second component is also chosen for this second node. If both
components of the second node are stubs, a third node with two free
stubs is selected and the triangle is closed. If one component of
the second node is a stub and another an edge, the node at the end
of the edge will be the third node and the triangle is formed linking
stubs between the first and the second node and the first and the
third node (see Fig. \ref{figD3}).
\begin{figure}[h]
\epsfig{file=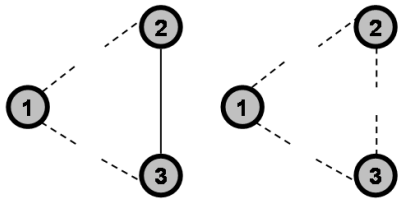,width=6cm}
 \caption{The two selected components of the first node,
  marked 1, are stubs.} \label{figD3}
\end{figure}

\item After each triangle is formed, all dynamic
quantities are updated: linked stubs are converted into edges and
the corresponding number of new triangles is added to all involved
degree classes. It is worth to mention that not only one more triangle is computed for the
classes of the nodes forming the triangle, but the degree classes of
simultaneous neighbors of pairs of those nodes may also be affected if those pairs were not previously connected.
The set EC is also updated, removing components of nodes in
new satisfied classes as well as nodes with only one component or
nodes which cannot form more triangles.
\item This process is repeated until all classes are satisfied or there are no more components in
the eligible set.
\end{itemize}

\noindent
{\bf 3) Closure of the network}\\

The final step consists in the closure of the network
by applying the classical configuration model to the remainder stubs.
Pairs of these stubs are selected uniformly at random and the
corresponding vertices are connected by an undirected edge.\\

In this way, the algorithm is able to reproduce networks with a given
degree distribution, $P(k)$, and a given clustering coefficient,
$\bar{c}(k)$, as long as the assortativity --that is, positive degree-degree correlations-- is high enough to avoid
constraining $\bar{c}(k)$. This is by no means a deficiency of the algorithm but an universal structural constraint imposed by the degree-degree correlation pattern of the network. In general, with the maximum
assortativity one can reproduce any desired level of clustering, whereas disassortative networks have instead a bounded clustering coefficient. A theoretical explanation is given in section V.

In our algorithm, the level of assortativity is controlled by the
probability by which the degree class is chosen previously to the
selection of the node. This can be done in a number of different ways. In our case, we tune the assortativity by
choosing a proper form for the probability $\Pi(k)$. For instance, an
uniparametric function modeling different assortativity levels is
given by $\Pi(k) \propto T_r(k)^\beta$, where $T_r(k)$ is the
number of triangles remaining to be formed in the degree class $k$
in a given iteration. The value of $\beta$ typically ranges in the interval $[0,1]$, generating more assortative networks as $\beta$ approaches $0$.
\begin{figure}
\epsfig{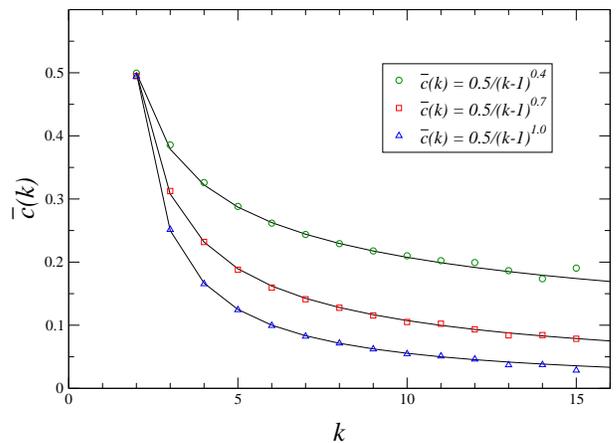}
 \caption{Clustering coefficient for networks generated by the algorithm using a Poisson degree distribution with average degree $\langle k \rangle =4$ and expected clustering coefficient $\bar{c}(k)=c_{0} (k-1)^{-\alpha}$ (solid lines) with $\alpha=1$, $0.7$, $0.4$, and $c_{0}=0.5$ in all cases. Each curve is an average over three different realizations with a network size of $N=10^5$. The parameter $\beta$ is equal to $1$ for $\alpha=1$ and $\alpha=0.7$, and $\beta=0.5$ for $\alpha=0.4$.} \label{fig6}
\end{figure}

\section{Numerical simulations}

To check the feasibility and reliability of the algorithm, we have performed extensive numerical simulations, generating networks with different types of degree distributions and different levels of clustering. The chosen forms for the degree distribuion are Poisson, exponential, and scale-free. The degree dependent clustering coefficient is chosen to be $\bar{c}(k)=c_{0} (k-1)^{-\alpha}$. The numerical prefactor is set to $c_{0}=0.5$ and the exponent takes values $\alpha=1$, $0.7$, and $0.4$. The size of the generated networks is $N=10^5$ and each curve is an average over three different realizations.

\begin{figure}
\epsfig{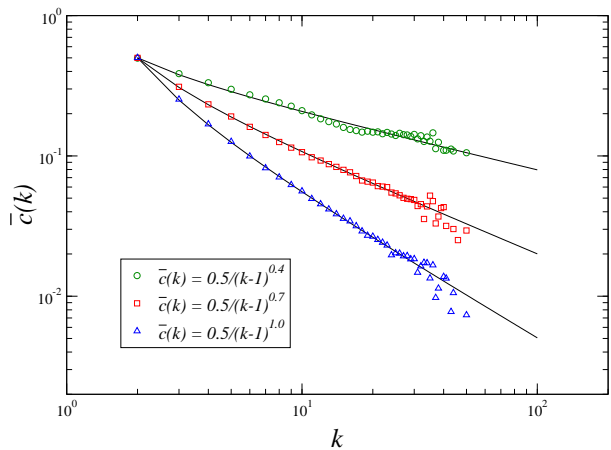}
 \caption{The same as in Fig.\ref{fig6} for an exponential degree distribution. In this case, the parameter $\beta$ is $1$ for $\alpha=1$ and $\alpha=0.7$, and $\beta=0$ for $\alpha=0.4$.} \label{fig7}
 \end{figure}

\begin{figure}
\epsfig{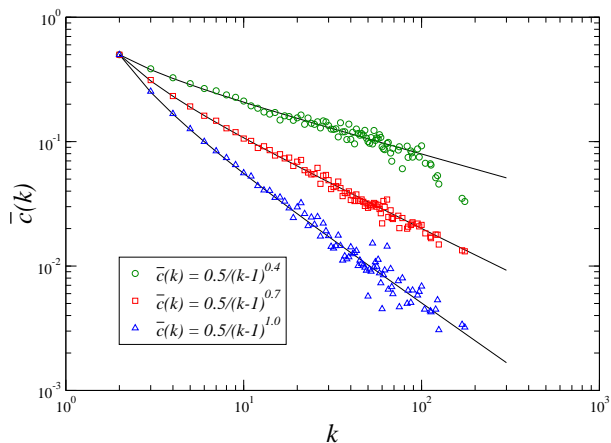}
 \caption{The same as in Fig. \ref{fig6} for an scale-free degree distribution of exponent $\gamma=3$. In this case, the parameter $\beta$ is $1$ for $\alpha=1$ and $\alpha=0.7$, and $\beta=0.2$ for $\alpha=0.4$.} \label{fig8}
\end{figure}

Simulation results are shown in Figs. \ref{fig6}, \ref{fig7}, and \ref{fig8}, which correspond to Poisson and exponential degree distribution with average degree $\langle k \rangle=4$, and scale-free degree distributions with exponent $\gamma=3$, respectively. As it can be seen, the degree dependent clustering coefficient is well reproduced in all cases just by decreasing the value of $\beta$ if necessary (the values of $\beta$ used in each simulation are specified in the caption of the corresponding figure). The standard procedure we follow is to start with $\beta=1$ and to check whether the tail of $\bar{c}(k)$ is well reproduced. If not, we decrease its value until the entire curve fits the expected shape.

\begin{figure}
\epsfig{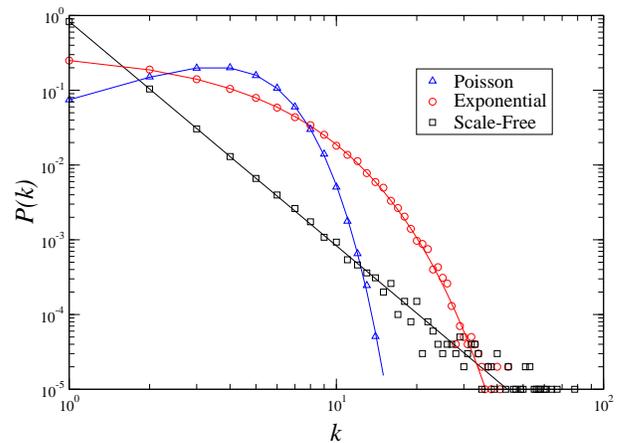}
 \caption{Degree distributions generated by the algorithm (symbols) as compared to the expected ones (solid lines).} \label{P_k}
\end{figure}

Fig. \ref{P_k}  shows the degree distributions generated by the algorithm for the simulations of the previous figures, confirming that, indeed, the generated degree distributions match the expected ones. 

\section{Clustering vs. degree-degree correlations}
\label{theoretical}
As we advanced in the previous section, degree-degree correlations constraint the maximum level of clustering a network can reach. A naive explanation for this is that, if the neighbors of a given node have all of them a small degree, the number of connected neighbors (and hence, the clustering of such node) will be bounded. This is the main idea behind the new measure of clustering introduced in \cite{ClusteringVazquez}. However, we can make a step forward and quantify analytically this effect. To do so, 
we need to define new quantities which take into account the
properties of vertices that belong to the same triangle. Let us
define the multiplicity of an edge, $m_{ij}$, as the number of
triangles in which the edge connecting vertices $i$ and $j$
participates. This quantity is the analog to the number of triangles
attached to a vertex, $T_{i}$. These two quantities are related
through the trivial identity
\begin{equation}
\sum_{j} m_{ij}a_{ij} = 2 T_i,
\end{equation}
which is valid for any network configuration. The matrix $a_{ij}$ is
the adjacency matrix, giving the value $1$ if there is an edge
between vertices $i$ and $j$ and $0$ otherwise. 

It is possible to
find a relation between multiplicity, degree distributions and
clustering. Summing the above equation for all vertices of a given
degree class we get
\begin{equation}
\sum_{k'} \sum_{i \in \Upsilon(k)} \sum_{j  \in \Upsilon(k')}
m_{ij}a_{ij} = \sum_{i \in \Upsilon(k)} 2 T_i. \label{eqorigen}
\end{equation}
Now, there are some key relations which can be used
\begin{equation}
\sum_{i \in \Upsilon(k)} \sum_{j  \in \Upsilon(k')} m_{ij}a_{ij} =
m_{kk'}E_{kk'},
\end{equation}
where $m_{kk'}$ is the average multiplicity of the edges connecting
the classes $k$ and $k'$, and $E_{kk'}$ is the number of edges
between those degree classes. Finally, taking into account Eq. (\ref{c(k)})
and the fact that the joint degree distribution satisfies
$P(k,k')= \lim_{N \rightarrow \infty} E_{kk'}/\langle k \rangle N$, we obtain the following
closure condition for the network
\begin{equation}
\sum_{k'} m_{kk'}P(k,k') = k(k-1) \bar{c}(k) \frac{P(k)}{\langle k
\rangle}. \label{db}
\end{equation}
Let us emphasize that this equation is, in fact, an identity fulfilled by any network and, thus, it is, for instance, at the same level as the degree detailed balance condition derived in \cite{Boguna02}. These identities are important because, given their universal nature, they can be used to derive properties of networks regardless their specific details. As an example, in \cite{Boguna03} we used the detailed balance condition to prove the divergence of the maximum eigenvalue of the connectivity matrix that rules the epidemic spreading in scale-free networks, which, in turn, implies the absence of epidemic threshold in this type of networks. 

The multiplicity matrix is, {\it per se}, a very interesting object that gives a more detailed description
on how triangles are shared by vertices of different degrees. In
principle, $m_{kk'}$ does not factorize and, therefore, non trivial correlations can be found. The global average
multiplicity of the network, $\bar{m}$, can be computed as
\begin{equation}
\bar{m}=\sum_{k}\sum_{k'} m_{kk'}P(k,k')=\frac{\langle k
(k-1)\bar{c}(k) \rangle}{\langle k \rangle}.
\end{equation}
Values of $\bar{m}$ close to zero mean that there are no triangles.
When $\bar{m} \approx 1$, triangles are mostly disjoint and their number can
be approximated as $T(k)\approx k/2$, and, when $\bar{m} \gg 1$,
triangles jam into the edges, that is, many triangles share common edges.

We are now equipped with the necessary tools to analyze the interplay between degree-degree correlations and clustering. The key point is to realize that the multiplicity matrix satisfies the inequality 
\begin{equation}
m_{kk'} \le \min(k,k')-1,
\end{equation}
which comes from the fact that the degrees of the nodes at the ends of an edge determine the maximum number of triangles this edge can hold. Multiplying this inequality by $P(k,k')$ and summing over $k'$ we get
\begin{equation}
k(k-1) \bar{c}(k) \frac{P(k)}{\langle k \rangle} \le \sum_{k'}
\min(k,k')P(k,k')-\frac{kP(k)}{\langle k \rangle},
\end{equation}
where we have used the identity Eq. (\ref{db}). This inequality, in
turn, can be rewritten as
\begin{equation}
\bar{c}(k) \le 1-\frac{1}{k-1} \sum_{k'=1}^{k} (k-k') P(k'|k) \equiv \lambda(k).
\label{inequality}
\end{equation}
Notice that $\lambda(k)$ is always in the interval $[0,1]$ and, therefore, $\bar{c}(k)$ is always bounded by a function smaller (or equal) than $1$. In the limit of very large values of $k$, Eq. (\ref{inequality}) reads
\begin{equation}
\bar{c}(k) \le \lambda(k) \approx \frac{\bar{k}_{nn}^{r}(k)-1}{k-1}
\label{inequality_approx}
\end{equation}
where $\bar{k}_{nn}^{r}(k)$ is the average nearest neighbors degree
of a vertex with degree $k$. The superscript $r$ (of reduced) refers to the fact
that it is evaluated only up to $k$ and, therefore,
$\bar{k}_{nn}^{r}(k) \le k$. For strongly assortative networks
$\bar{k}_{nn}^{r}(k) \sim k$, so that $\lambda(k) \sim {\cal O}(1)$ and there is no restriction in
the decay of $\bar{c}(k)$. In the opposite case of disassortative networks, the sum term in the
right hand side of Eq. (\ref{inequality}) may be fairly large and then the clustering coefficient will have to decay
accordingly. 

\begin{figure}[h]
\epsfig{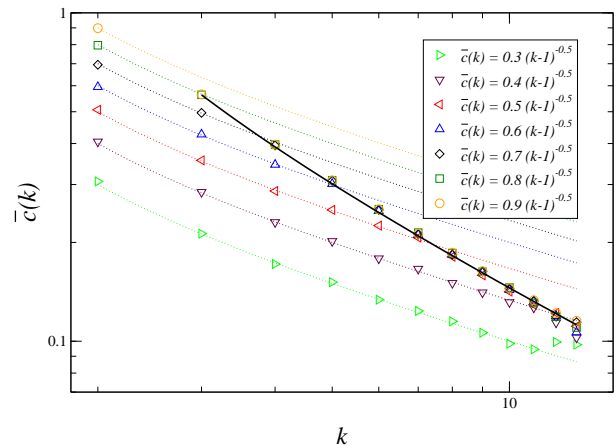}
 \caption{Clustering coefficient for Poisson like degree distributions, using $\beta=1$. Different
 curves correspond to different values of the prefactor $c_{0}$. Doted lines are the expected clusterings whereas symbols are the ones generated by the algorithm. The solid line is a guide for the eye of the limiting curve. For lower values of the
 prefactor, the expected value can be fitted in a wider region. Notice that all curves
 collapse into the same limiting curve, which indicates the intrinsic constraint Eq.(\ref{inequality}).} \label{fig1}
\end{figure}

In Fig.\ref{fig1} we show this effect by changing the level of
clustering while keeping the degree-degree correlations unchanged by fixing the value of $\beta$ to $\beta=1$.
As it can be seen, lower levels of clustering are better reproduced.
However, the clustering collapses to a limiting curve when the
expected value crosses it. That is, any function $\bar{c}(k)$ is possible whenever it is defined below a limiting curve which is a function of the degree correlation pattern of the network.

\begin{figure}[h]
\epsfig{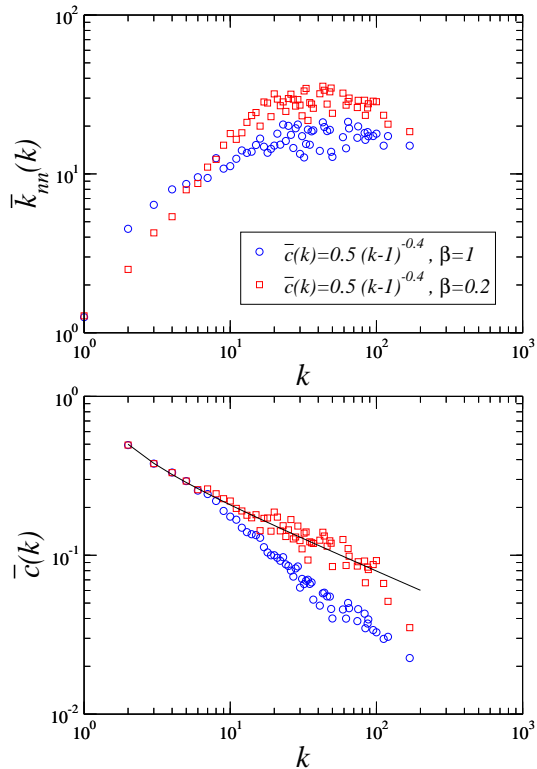}
 \caption{Average nearest neighbors degree (top), $\bar{k}_{nn}(k)$, and clustering coefficient (bottom), $\bar{c}(k)$,
 for a power law degree distribution with exponent $\gamma=3$ using two different levels of assortativity, $\beta=1$ and $\beta=0.2$. As we increase assortativity the expected clustering can be fitted in a wider region. The solid line is the expected clustering $\bar{c}(k)=0.5 (k-1)^{-0.4}$.} \label{fig2}
\end{figure}

Another way to see the same effect is shown in Fig. \ref{fig2}. In this case we keep the expected clustering while changing the assortativity of the network by tuning the parameter $\beta$. As it can be seen, as correlations become more and
more assortative (decreasing values of $\beta$) the expected clustering can be further reproduced.

We would like to point out that the function $\lambda(k)$ is just an upper bound for the clustering coefficient. The actual bound will probably be even smaller due to the fact that we have only considered the restriction over
one edge and the degrees of the corresponding vertices. A more
accurate estimation would involve more than one edge and the
corresponding vertices attached to them \cite{ClusteringVazquez}.

\begin{figure}[h]
\epsfig{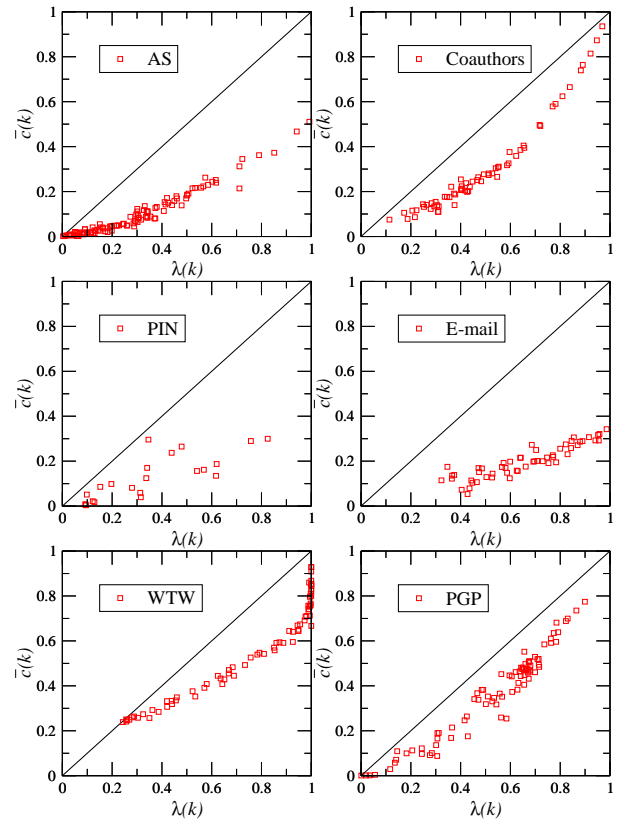}
 \caption{Clustering $\bar{c}(k)$ versus the maximum value $\lambda(k)$ for several real networks. In all cases, empirical measures fall below the diagonal line, validating the inequality Eq. (\ref{inequality}).} \label{inequality2}
\end{figure}

\subsection{Scale-free networks}

Scale-free networks belong to a special class of networks which deserve a separate discussion. Indeed, it has been shown that, when the exponent of the degree distribution lies in the interval $\gamma \in (2,3]$ and its domain extends beyond values that scale as $N^{1/2}$, disassortative correlations are unavoidable for high degrees \cite{Park03,Burda03,Boguna04a,Catanzaro05}. Almost all real scale-free networks fulfill these conditions and, hence, it is important to analyze how these negative correlations constraint the behavior of the clustering coefficient. Let us assume a power law decay of the average nearest neighbors degree of the form $\bar{k}_{nn}(k)\sim \kappa k^{-\delta}$. One can prove that this function diverges in the limit of very large networks as $\bar{k}_{nn}(k)\sim \langle k^2 \rangle \sim k_{c}^{3-\gamma}$, where $k_{c}$ is the maximum degree of the network \cite{Boguna03}. Then, the prefactor $\kappa$ must scale in the same way which, in turn, implies that the reduced average nearest neighbors degree behaves as
\begin{equation}
\bar{k}_{nn}^r(k)\sim  k^{3-\gamma-\delta}.
\end{equation}
Then, from Eq. (\ref{inequality_approx}) the exponent of the degree dependent clustering coefficient, $\alpha$, must verify the following inequality
\begin{equation} 
\alpha \ge \gamma+\delta-2.
\end{equation}
Just as an example, in the case of the Internet at the Autonomous System level \cite{DefclusteringkVespas}, the reported values for these three exponents ($\alpha=0.75$, $\gamma=2.1$, and $\delta=0.5$) satisfy this inequality close to the limit ($\alpha=0.75 \ge \gamma+\delta-2=0.6$).

\subsection{Real networks}
\begin{figure}[h]
\epsfig{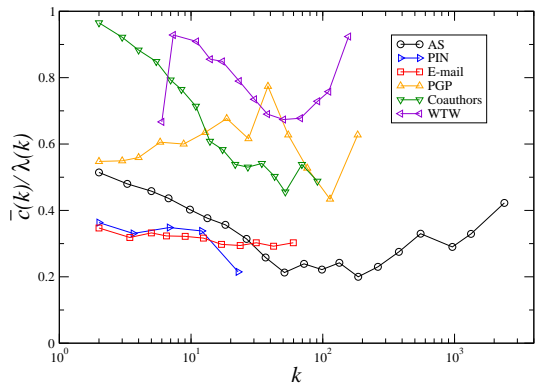}
 \caption{Empirical measures of the ratio between the clustering coefficient $\bar{c}(k)$ and the maximum value $\lambda(k)$ for different real networks.} \label{constraint_realnets}
\end{figure}

The interplay between degree correlations and clustering can also be observed in real networks. We have measured the functions $\lambda(k)$ and $\bar{c}(k)$ for several empirical data sets, finding that the inequality Eq.~(\ref{inequality}) is always satisfied. The analyzed networks are the Internet at the Autonomous System level (AS) \cite{Romusbook}, the protein interaction network of the yeast {\it S. Cerevisiae} (PIN) \cite{Jeong01}, an intra-university e-mail network \cite{Guimera03}, the web of trust of PGP \cite{Boguna04}, the network of coauthorships among academics \cite{Newman01}, and the world trade web (WTW) of trade relationships among countries \cite{Serrano03}. 

In Fig.~\ref{inequality2} we plot the clustering coefficient $\bar{c}(k)$ as a function of $\lambda(k)$. Each dot in these figures correspond to a different degree class. As clearly seen, in all cases the empirical measures lie below the diagonal line, which indicates that the inequality Eq.~(\ref{inequality}) is always preserved. In Fig.~\ref{constraint_realnets} we show the ratio $\bar{c}(k)/\lambda(k)$. The rate of variation of this fraction is small and, thus, the degree dependent clustering coefficient can be computed as $\bar{c}(k)=\lambda(k) f(k)$, where $f(k)$ is a slowly varying function of $k$ that, in many cases, can be fitted by a logarithmic function.

\section{Conclusions}

We have introduced and tested a new algorithm that generates {\it ad hoc} clustered networks with given degree distribution and degree dependent clustering coefficient. This algorithm will be useful for analyzing, in a controlled way, the role that clustering has on many dynamical processes that take place on top of networks. We have also introduced a new formalism which backs our algorithm and allows to quantify clustering in a more rigorous manner. In particular, an universal closure condition for networks is found to relate the degree dependent clustering coefficient, degree-degree correlations and the number of triangles passing through edges connecting vertices of different degree classes. Using this relation, we have found how the correlation pattern of the network constraints the function $\bar{c}(k)$. In particular, assortative networks are allowed to have high levels of clustering whereas disassortative ones are more limited. Overall, we hope that a more accurate shaping of synthetic networks will improve our understanding of real ones. At this respect, we believe our algorithm will be useful for the community working on complex networks science.

\begin{acknowledgments}
We acknowledge A. Vespignani, R. Pastor-Satorras and A. Arenas for valuable suggestions. This work
has been partially supported by DGES of the Spanish government,
Grant No. FIS2004-05923-CO2-02, and EC-FET Open project COSIN
IST-2001-33555. M. B.  acknowledges financial support from the
MCyT (Spain) through its Ram\'on y Cajal program.
\end{acknowledgments}

\end{document}